\documentclass[twocolumn,showpacs,preprintnumbers,aps,pra,amsmath,amssymb,floatfix,superscriptaddress]{revtex4-1}

\usepackage{graphicx}
\usepackage{dcolumn}
\usepackage{bm}
\usepackage{epsfig}
\usepackage{color}
\usepackage{natbib}

\newcommand{\ket}[1]{\left|{#1}\right\rangle}

\begin{document}

\title{Long-range states of the NaRb molecule near the Na($3^2S_{1/2}$)+Rb($5^2P_{3/2}$) asymptote}
\author{Bing Zhu}
\affiliation{Department of Physics, The Chinese University of Hong Kong, Hong Kong, China}
\author{Xiaoke Li}
\affiliation{Department of Physics, The Chinese University of Hong Kong, Hong Kong, China}
\author{Xiaodong He}
\altaffiliation[Permanent Address: ]{Wuhan Institute of Physics and Mathematics, Chinese Academy of Sciences, Wuhan 430071, China}
\author{Mingyang Guo}
\affiliation{Department of Physics, The Chinese University of Hong Kong, Hong Kong, China}
\author{Fudong Wang}
\affiliation{Department of Physics, The Chinese University of Hong Kong, Hong Kong, China}
\author{Romain Vexiau}
\affiliation{Laboratoire Aim\'e Cotton, CNRS, Universit\'e Paris-Sud, ENS Cachan, Universit\'e Paris-Saclay, 91405 Orsay Cedex, France}
\author{Nadia Bouloufa-Maafa}
\affiliation{Laboratoire Aim\'e Cotton, CNRS, Universit\'e Paris-Sud, ENS Cachan, Universit\'e Paris-Saclay, 91405 Orsay Cedex, France}
\author{Olivier Dulieu}
\affiliation{Laboratoire Aim\'e Cotton, CNRS, Universit\'e Paris-Sud, ENS Cachan, Universit\'e Paris-Saclay, 91405 Orsay Cedex, France}
\author{Dajun Wang}
\email{djwang@phy.cuhk.edu.hk}
\affiliation{Department of Physics, The Chinese University of Hong Kong, Hong Kong, China}
\affiliation{The Chinese University of Hong Kong Shenzhen Research Institute, Shenzhen, China} 

\date{\today}

\begin{abstract}

We report a high-resolution spectroscopic investigation of the long-range states of the $^{23}$Na$^{87}$Rb molecule near its Na($3^2S_{1/2}$)+Rb($5^2P_{3/2}$) asymptote. This study was performed with weakly bound ultracold molecules produced via magneto-association with an inter-species Feshbach resonance. We observed several regular vibrational series, which are assigned to the 5 attractive long-range states correlated with this asymptote. The vibrational levels of two of these states have sharp but complex structures due to hyperfine and Zeeman interactions. For the other states, we observed significant linewidth broadenings due to strong predissociation caused by spin-orbit couplings with states correlated to the lower Na($3^2S_{1/2}$)+Rb($5^2P_{1/2}$) asymptote. The long-range $C_6$ van der Waals coefficients extracted from our spectrum are in good agreement with theoretical values.

\end{abstract}

\maketitle


\section{Introduction}
Due to their promising applications in a wide range of areas, ultracold ground-state polar molecules with high phase-space density have been pursued intensively in the last two decades~\cite{EPJD2004, Carr2009}. Although recent years have seen significant progress in direct laser cooling of ground-state molecules~\cite{Shuman2010,Hummon2013, Barry2014}, the most successful way of producing near-quantum-degeneracy molecular samples is still association of ultracold atoms with the help of magnetic Feshbach resonances~\cite{Kohler06}. With a properly designed two-photon stimulated Raman process, the weakly-bound molecules formed can be transferred to the lowest energy level of the molecular ground state with the high initial phase-space density preserved~\cite{Winkler2006,Danzl2008,Ospelkaus2008}. This was successfully implemented for $^{40}$K$^{87}$Rb molecule~\cite{Ni2008} in 2008, and very recently for $^{87}$Rb$^{133}$Cs~\cite{Takekoshi2014,Molony2014} and $^{23}$Na$^{40}$K~\cite{Park2015} molecules. For $^{40}$K$^{87}$Rb, two-body chemical reaction can happen, which induces large inelastic losses~\cite{Ospelkaus10,Ni10}. $^{87}$Rb$^{133}$Cs and $^{23}$Na$^{40}$K are stable against two-body chemical reactions~\cite{Zuchowski2010}, but other loss mechanisms may still exist~\cite{Mayle2013,Takekoshi2014,Park2015}.

The bosonic $^{23}$Na$^{87}$Rb molecule in its ground state is also chemically stable against two-body collisions~\cite{Zuchowski2010}. Besides, it possesses a permanent electric dipole moment of 3.3 Debye, larger than all the aforementioned molecules~\cite{Aymar2005}. It is thus another important candidate for investigating ultracold polar molecules. In the past two years, we have prepared an ultracold mixture of $^{23}$Na and $^{87}$Rb~\cite{Xiong2013} and studied their Feshbach resonances~\cite{Wangfudong2013}. Recently, we have successfully produced $^{23}$Na$^{87}$Rb Feshbach molecules by sweeping the magnetic field across a Feshbach resonance~\cite{Wangfudong2015}.

\begin{figure}[hbtp]
\includegraphics[width=0.85 \linewidth]{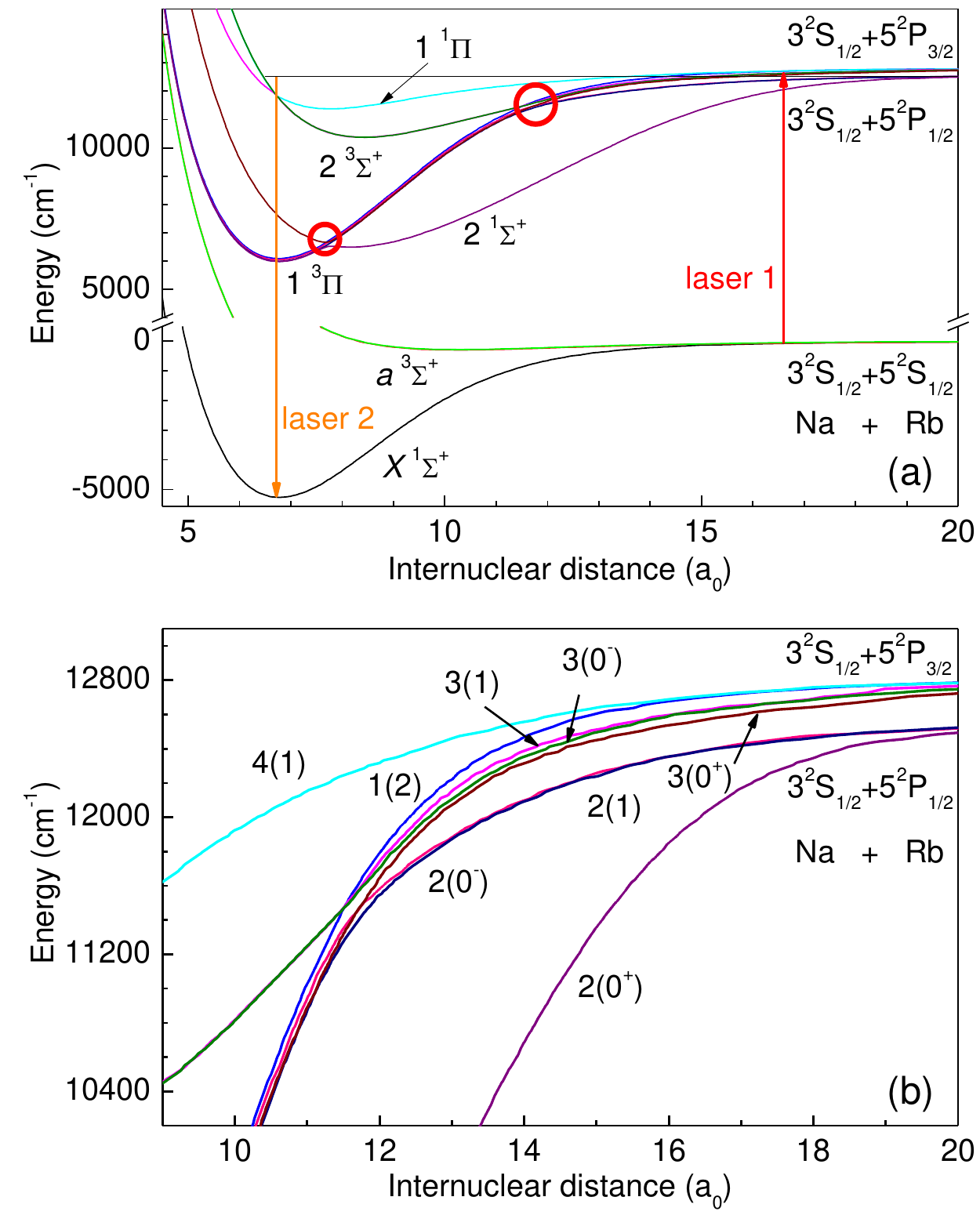}
\caption{\label{fig1}(color online). 
NaRb potential energy curves including spin-orbit coupling obtained from quantum chemistry computations~\cite{Korek2009}. (a) The long-range part of the excited potential can be accessed with laser 1 starting from weakly bound Feshbach molecules. A possible two-photon population transfer path which can be completed by adding laser 2, is also shown schematically. (b) The eight attractive long-range potentials correlated with Na($3^{2}S_{1/2}$)+Rb($5^{2}P$)  asymptotes are shown in a zoomed in view. The current work focuses on the 5 states correlated with the Na($3^{2}S_{1/2}$)+Rb($5^{2}P_{3/2}$) asymptote. Predissociation can happen for some of these states due to the curve crossings with states from the lower asymptote, near the region marked by the two red circles in (a). All internuclear distances are in the unit of Bohr radius ($a_0$). }
\end{figure}

The next critical step for making ground-state $^{23}$Na$^{87}$Rb molecules is to find a path for an efficient stimulated Raman process. As illustrated in Fig.\ref{fig1}, the internuclear distances between the ground-state and the Feshbach molecules are vastly different. Besides, as our Feshbach molecules are mainly of triplet character, the intermediate level must have strong triplet/singlet mixing to make the connection to the $X^1\Sigma^+$ state. Thus, for this task, a detailed understanding of the excited states is necessary. Historically, $^{23}$Na$^{87}$Rb has been a popular molecule for spectroscopic study. The two electronic ground states, $X^1\Sigma^+$ and $a^3\Sigma^+$, were investigated in detail with Fourier transform spectroscopy~\cite{Pashov2005} and further refined by measurement of the Feshbach resonances~\cite{Wangfudong2013}. Several low-lying excited states shown in Fig.\ref{fig1}, including $1^1\Pi$~\cite{Wangyouchang1991,Pashov2006}, $2^1\Sigma^+$ and $1^3\Pi$~\cite{Docenko2007} were also well studied by various conventional spectroscopy methods.

However, as is typical for conventional molecular spectroscopy, the long-range parts of the excited state potentials are not covered well. In this work, we study the long-range states near the Na($3^{2}S_{1/2}$)+Rb($5^{2}P_{3/2}$) asymptote starting from Feshbach molecules which are weakly bound with a large amplitude vibrational motion and is thus a nice choice for this purpose. Besides its importance in understanding fundamental molecular dynamics, this work is also an important step toward finding the intermediate level for Raman transfer. One of the primary choices satisfying our requirements is the strongly mixed levels of the $1^1\Pi$ and $2^3\Sigma^+$ admixture~\cite{Ni2008,Borsalino2014}. As shown schematically in Fig.~\ref{fig1}(a), the equilibrium distance of the $X^1\Sigma^+$ state matches well with the classical inner turning points of some high vibrational levels of the $1^1\Pi$ and $2^3\Sigma^+$ mixture; thus favorable strengths are expected for both the pump and dump transitions.

This paper is organized as follows: In Sec.~\ref{experiment}, we describe our Feshbach molecule preparation system and the spectroscopy setup. In Sec.~\ref{result} and Sec.~\ref{discussion}, we show the high resolution spectrum and our analysis. Sec.~\ref{conclusion} concludes the article.

\section{Experiments}
 \label{experiment}

The experiment reported here is carried out on the same setup as our previous works~\cite{Xiong2013,Wangfudong2013,Wangfudong2015}. After preparing the ultracold mixture of Na and Rb atoms in their $\ket{F = 1, m_F=1}$ hyperfine states in an optical dipole trap, we perform magneto-association by sweeping the magnetic field across the inter-species Feshbach resonance located at 347.8 G. After removing residual atoms, we typically end up with about 1500 Feshbach molecules with a temperature of 450 nK, which is the starting point for the current work. The final magnetic field is at 340 G, where the molecules have a binding energy of $2\pi\times 12$ MHz and about 80$\%$ closed channel fraction. The trap lifetime of the pure molecule sample is more than 20 ms~\cite{Wangfudong2015}. 

The spectroscopic laser is a free-running external cavity diode laser with a maximum power of 50 mW and a typical linewidth of 2 MHz. Its frequency is measured by a wavelength meter (HighFinesse, WS7) with an absolute accuracy of 60 MHz. The laser beam propagates perpendicular to the magnetic field and is focused to a waist of $\sim$45 $\mu$m. As the transition strength varies with the binding energy, we adjust the laser power to maintain similar loss ratios over the whole scan range. For the strongest transitions, $<$ 100 $\mu$w power is enough to deplete all the Feshbach molecules. The typical excitation duration is 100 $\mu$s, much shorter than the trap lifetime. 


We obtain the bound-bound spectrum by tuning the spectroscopic laser from 12816.5 cm$^{-1}$ to 12806 cm$^{-1}$ with typical steps of 50 MHz. After the excitation, we ramp the magnetic field back to dissociate the remaining Feshbach molecules, as our absorption detection is only sensitive to atoms. We have to prepare a new sample after each cycle, which takes about 45 seconds. 

Our detuning measurements are all made with respect to the transition from the ground-state hyperfine asymptote $^{23}$Na(3$^2S_{1/2}$, $\ket{F=1,m_F=1}$) + $^{87}$Rb(5$^2S_{1/2}$, $\ket{F=1,m_F=1}$) to the center of gravity of the Na($3^{2}S_{1/2}$)+Rb($5^{2}P_{3/2}$) asymptote. After taking the Zeeman shifts at 340 G into account, the excited-state asymptotic limit is taken as 12816.724 cm$^{-1}$. The small binding energy of the Feshbach molecules is neglected.

\section{Spectrum and assignments}
 \label{result} 

\begin{figure*}[hbtp]
\centering\includegraphics[width=0.9\linewidth]{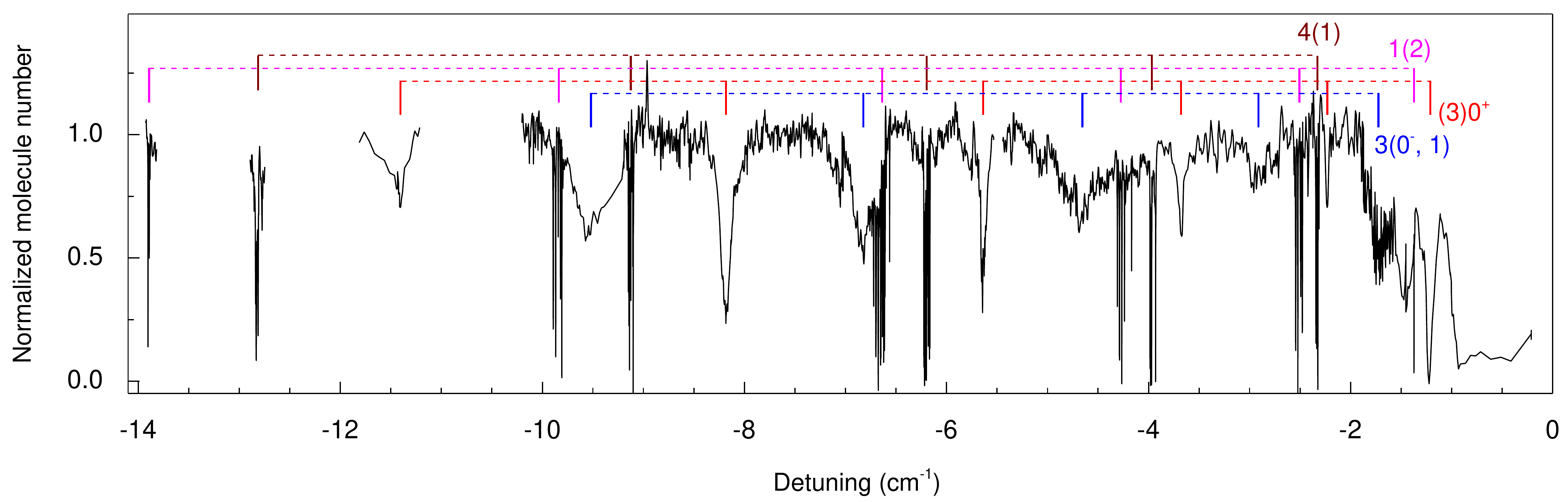}
\caption{(color online) A long spectrum shows all observed excited-state levels. 4 vibrational series are distinguished from each other by their very different lineshapes and distributions. Each series contains 5 to 6 levels. Our assignments are depicted by 4 sets of vertically displaced solid bars. The bottom dashed line correlates with \textit{both} 3(1) and 3(0$^-$).}
\label{fig2}
\end{figure*}  

In Fig.~\ref{fig2} we show the whole spectrum near the Na($3^{2}S_{1/2}$)+Rb($5^{2}P_{3/2}$) asymptote. Levels with detuning less than -1.2 cm$^{-1}$ are not resolved, probably due to the very dense and strong resonances near this region, as well as predissociation-induced broadening which we will discuss below. At a first glance, the spectrum seems rather complex, reflecting the fact that there are 5 attractive long-range potentials, $3(0^{+})$, $3(0^{-})$, $3(1)$, $4(1)$ and $1(2)$, associated with this asymptote. Here we are using the Hund's case (c) notation typical for such long-range molecules, with the number in the parenthesis representing $\Omega$, the projection of the total angular momentum of the electrons on the molecular axis, and the first number their order with increasing energy in this symmetry.  

\begin{figure}[hbtp]
\includegraphics[width=0.9 \linewidth]{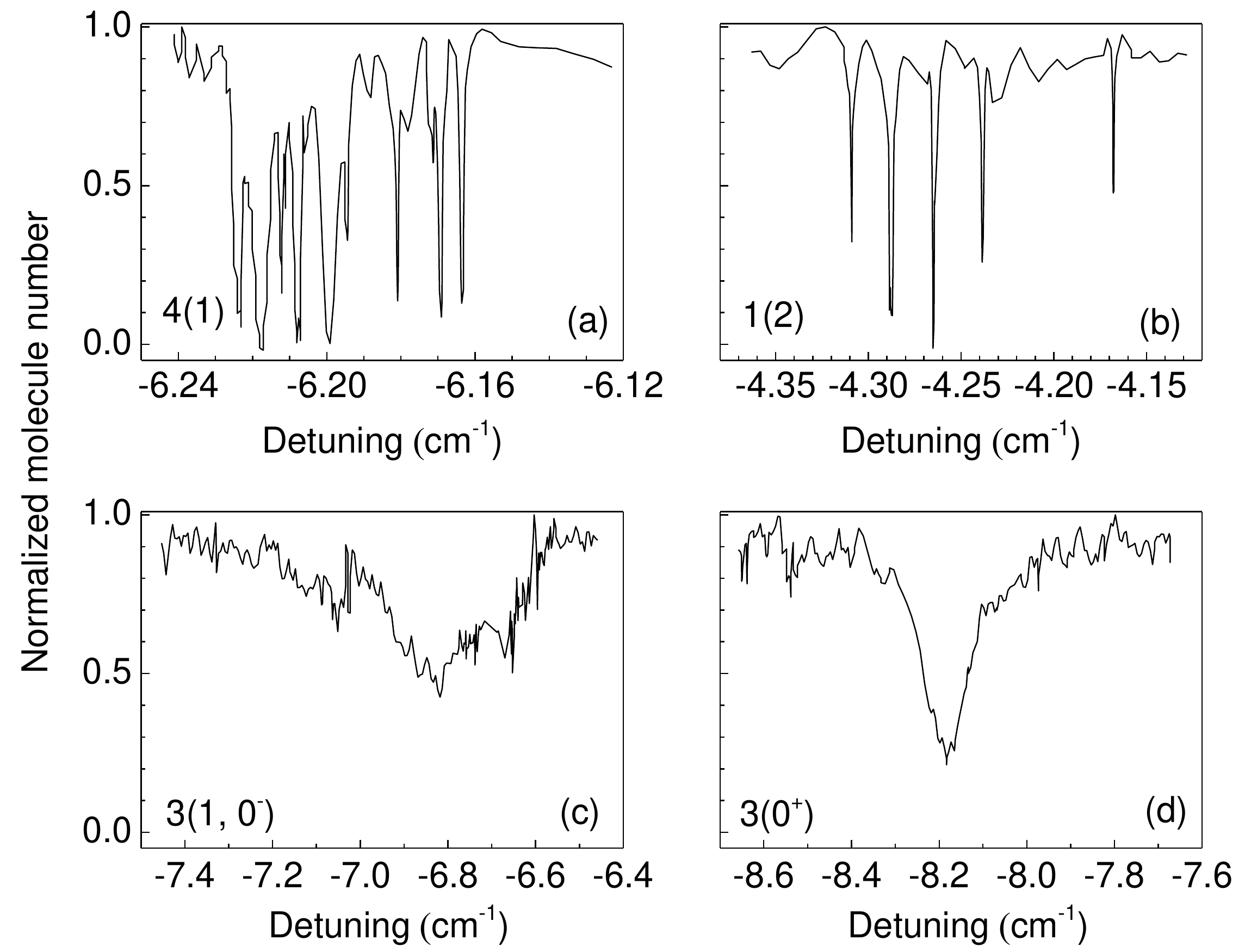}
\caption{\label{fig3} 
High resolution spectrum of different states. (a), (b), (c) and (d) show typical lineshapes of 4(1), 1(2), 3(1,0$^{-}$) and 3(0$^+$) states, respectively. The sharp structures in 4(1) and 1(2) levels are from hyperfine and Zeeman interactions. The very broad linewidths in 3(1,0$^{-}$) and 3(0$^+$) levels are due to fast predissociation. See text for details.}
\end{figure}

Scrutinizing the lineshape patterns and level spacings, we have picked out 4 vibrational series from the spectrum and listed them in Table~\ref{table1}. One key to achieve sensible assignments is the well resolved hyperfine and Zeeman structures for $\Omega > 0$ states. As shown in Fig.~\ref{fig3}(a) and (b), vibrational levels in two of the identified series have sharp sub-structures. To tell which state they belong to, we compare our spectrum to calculated vibrational levels based on the Rydberg-Klein-Rees (RKR) potential of $1^1\Pi$ state given by Pashov et al. in ref.~\cite{Pashov2006}. The 4(1) state is correlated to the $1^1\Pi$ state at long range. Good matching to the series including the level in Fig.~\ref{fig3}a is found by an overall -0.15 cm$^{-1}$ shift of the calculated vibrational levels. This series is thus assigned to the 4(1) state. 

\begin{table}
\caption{
Summary of the observed long-range vibrational levels near the Na($3^{2}S_{1/2}$)+Rb($5^{2}P_{3/2}$) asymptote. The $v$ index is deduced from a LeRoy-Bernstein fit (see text). Note that $v = -1$ represents the highest vibrational level. The linewidths of levels with predissociation are also listed.     
}
\begin{tabular}{c|c|c|c}
\hline
  state & $E_{v}~\rm{(cm^{-1})}$ & $v$ & linewidth (GHz) \\
\hline
  4(1) & 12814.3843 & -6 &  - \\
  		 & 12812.7455 & -7 &  - \\
 			 & 12810.5167 & -8 &  - \\
 		   & 12807.5884 & -9 &  - \\
 		   & 12803.8984 & -10 &  - \\
\hline
  1(2) & 12815.3403 & -5 &  - \\
  		 & 12814.2063 & -6 &  - \\
  		 & 12812.4381 & -7 &  - \\
  		 & 12810.0764 & -8 &  - \\
  		 & 12806.8752 & -9 &  - \\
  		 & 12802.8207 & -10 &   - \\
\hline
  3(1, 0$^-$)	& 12814.9872 & -6 & 9.4  \\
  						& 12813.8007 & -7 & 8  \\
   					  & 12812.0585 & -8 & 10  \\
   						& 12809.8893 & -9 & 10  \\
   						& 12807.194 & -10 &  - \\
\hline
  3(0$^+$) & 12815.5017 & -5 & 3  \\
  				 & 12814.4814 & -6 & 3.4  \\
  				 & 12813.0347 & -7 & 1.8  \\
  				 & 12811.0757 & -8 & 1.7  \\
  				 & 12808.5312 & -9 & 1  \\
  				 & 12805.3087 & -10 & 1.9  \\
\hline
\end{tabular}
  \label{table1}
\end{table} 

Another key for the assignment is the predissociation caused by coupling with long-range states in the lower Na($3^{2}S_{1/2}$)+Rb($5^{2}P_{1/2}$) asymptote. A similar mechanism, which typically results in significant linewidth broadening, has been observed in several other alkali dimers~\cite{Bergeman2006}. It affects the $3(0^{+})$, $3(0^{-})$ and $3(1)$ states, as can be seen from the long-range potential in Fig.~\ref{fig1}(b). The $4(1)$ and $1(2)$ states, on the other hand, are free from predissociation. Thus the vibrational series including the level in Fig.~\ref{fig3}b can only come from the $1(2)$ state, as the well resolved sub-structures with sharp peaks indicate no sign of predissociation. Detailed assignment of these sub-structures is still lacking at this point.

Our Feshbach molecule is s-wave in nature; thus its pure rotation is $l = 0$. It is electronic spin triplet dominated with the second to the last vibrational level ($v''=-2$) of the $a^3\Sigma^+$ state as the closed channel~\cite{Wangfudong2015}. The total electronic angular momentum is thus $j = 1$. Ignoring nuclear spins, the total angular momentum of the molecule ($\vec{J} = \vec{l}+\vec{j}$) can then only be $J''=1$. The selection rules of $J$ allow transitions with $\Delta J = 0$ and $\pm 1$, depending on $\Omega$ of the excited state. As the rotational levels of $\Omega = 2$ states start from $J' = 2$, the observation of  the $1(2)$ state proves the total angular momentum assignment of the Feshbach molecule is correct. The s-wave Feshbach molecule also has a + parity, and thus can only be promoted to excited states with - parity. All rotational levels of the $\Omega =$ 1 and 2 states have both + and - components due to the $\Lambda$-type doubling. Thus for $3(1)$ and $4(1)$ states, we can access $J' =$ 1 and 2, while for the $1(2)$ state, only $J' =$ 2 is possible. For the $0^+$ state, the electronic parity is even. Only odd $J'$ level can have - total parity, so only $J'=$ 1 can be accessed. While for the $0^-$ state, the electronic parity is odd; thus both $J'=$ 0 and 2 can be reached.

The predissociation observed in the other 3 states, however, obscures all rotational structure. As can be seen in Table~\ref{table1}, for the series including Fig.~\ref{fig3}(c), the extracted linewidths are on the order 10 GHz, while the other series have typical linewidths of 1 to 3 GHz. Typical rotational spacings for these near dissociation levels are estimated to be about 1 GHz, comparable or smaller than the observed linewidths. All of these values are much larger than the radiative lifetime limited linewidths, which are on the order of several MHz. We have verified experimentally that the contribution from power broadening is negligible. We note that even for the 4(1) state without predissociation, rotational levels are still hard to identify due to the comparable splittings from hyperfine and Zeeman interactions.  

To make the final assignment, we resort to the detailed study of the predissociation processes. As depicted in Fig.\ref{fig1}(a), the 3(0$^+$) state results from the spin-orbit coupling between the $A~^1\Sigma^+$ and the $b~^3\Pi$ states with a crossing at internuclear distance of about 7.7 $a_0$. The 3(0$^-$) and the main characters of the 3(1) states are from the spin-orbit coupling between the $c~^3\Sigma^+$ and the $b~^3\Pi$ states with a crossing at around 11.8 $a_0$. The linewidth can be estimated with a Landau-Zener model which takes into account the spin-orbit coupling strength between the adiabatic potential curves and the difference between their slopes at their crossing point. With this model, we obtain linewidth upper bounds of 2$\pi\times$2.7, 9.4 and 12.3 GHz for the 3(0$^+$), 3(1) and 3(0$^-$) states, respectively. Thus most probably, the series including Fig.~\ref{fig3}d is from 3(0$^+$) state. We attribute the series including Fig.~\ref{fig3}c to both 3(1) and 3(0$^-$), as those two states have very similar long-range dispersion coefficients and are expected to have closely spaced vibrational levels. With the very large predissociation linewidth, it becomes hard to resolve them individually.

\section{Long-range fitting and discussion}
 \label{discussion}

\begin{figure}[hbtp]
\includegraphics[width=0.9 \linewidth]{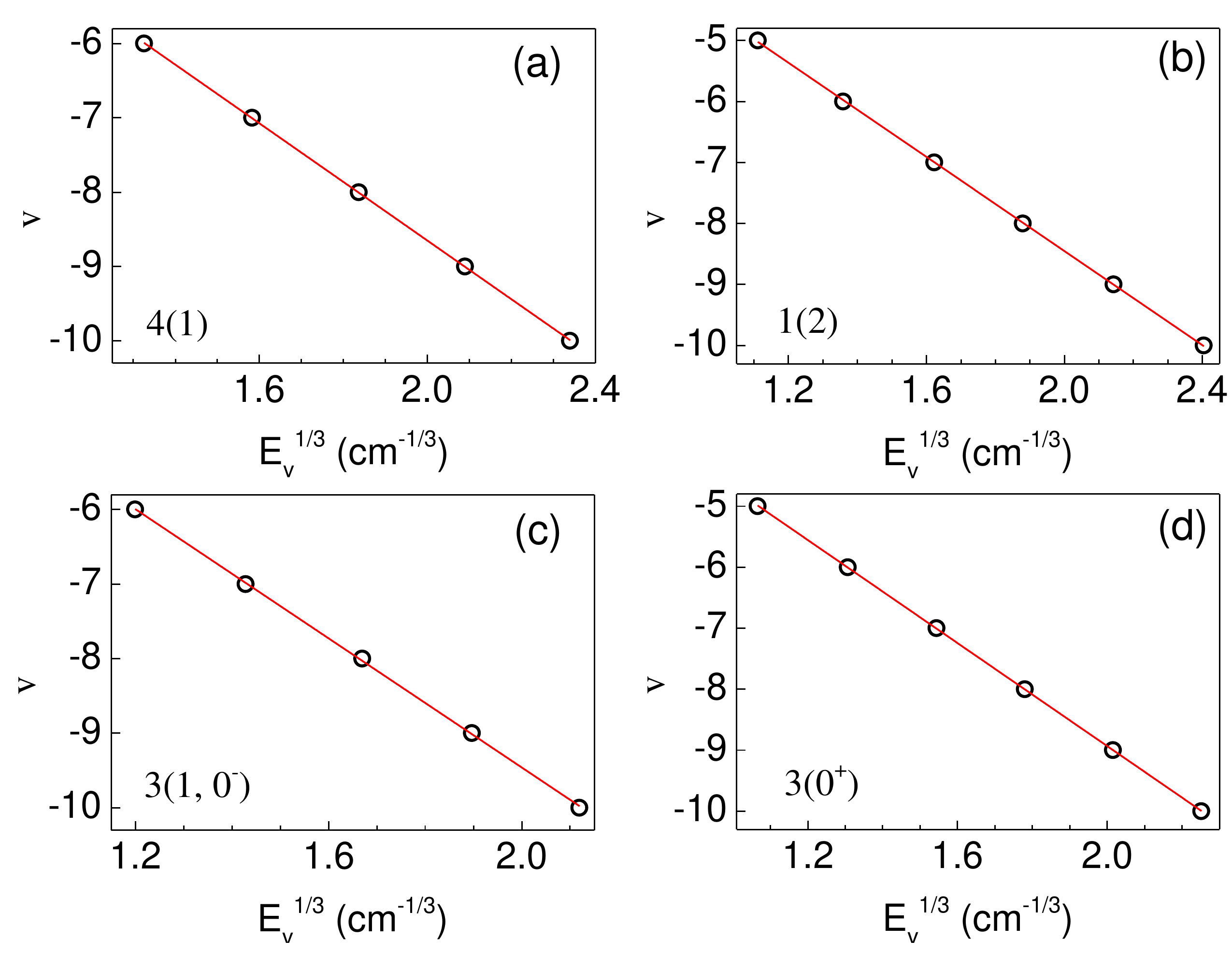}
\caption{\label{fig4}(color online). 
Plot of the the relative vibrational quantum number $v$ vs. 1/3 power of the vibrational binding energy E$_v$. The states in (a), (b), (c) and (d) are in the same order as Fig.~\ref{fig3}. The $C_6$ coefficients can be obtained from the slopes of the linear fittings. }
\end{figure}

With the assignment decided, the C$_{6}$ coefficients can be obtained by fitting the measured vibrational binding energies of each series to the semi-classical Leroy-Bernstein (LB) formula~\cite{LeRoy1970,Stwalley1970}
\begin{equation}
v-v_D=k_6 E_v^{1/3},
\end{equation}
with $k_{6}=\frac{2\sqrt{\pi}}{4h}\frac{\Gamma(3/2)}{\Gamma(7/6)}\sqrt{2\mu}C_{6}^{1/6}$. Here $v_{D}$ is the effective vibrational quantum number at dissociation (a number between -1 and 0 as we take the last bound level as $v$ = -1), $E_v$ is the binding energy, $h$ is the Planck constant, $\Gamma$ is the Gamma function, and $\mu$ is the reduced mass for $^{23}$Na$^{87}$Rb. Fig.~\ref{fig4} shows results of the $E_v^{1/3}$ vs. $v$ fittings for all the observed vibrational series. For each state, the $C_6$ coefficients are extracted from the fitting and listed in Table~\ref{table2}, together with the $C_6$ values from two previous calculations~\cite{Bussery1986,Marinescu1999}. 

The $C_6$ values of columns (a) and (b) in Table~\ref{table2} are deduced from the van der Waals coefficients for Hund's case (a) states calculated in Refs.~\cite{Marinescu1999} and~\cite{Bussery1986}, respectively, using the analytical formulas of Ref.~\cite{Frecon1998} which are exactly-solvable with the exchange energy neglected. The agreement between our results and the values deduced from Ref.~\cite{Marinescu1999} is quite reasonable, with less than 10$\%$  discrepancies for most of the states. The only exception is the 3(0$^+$) state, which shows a 16$\%$ difference. On the other hand, we observe a disagreement larger than 20$\%$ for most states with the values analytically deduced from Ref.~\cite{Bussery1986}. Such differences originate in those observed in the Hund's case (a) $C_6$ values between Refs.~\cite{Bussery1986} and~\cite{Marinescu1999}, related to the details of the computed atomic wave functions. For completeness, we also display in column (c) of Table~\ref{table2} the $C_6$ values reported in Ref.~\cite{Bussery1986}, where the authors fitted their numerical calculations.

\begin{table}[h]
\caption{$C_6$ coefficients from the LB fit and their comparison with several theoretical works. (a) and (b): from Ref.~\cite{Marinescu1999} and Ref.~\cite{Bussery1986} respectively, using the analytical formula of Ref.~\cite{Frecon1998}; (c) from Ref.~\cite{Bussery1986}.    
}
\begin{tabular}{l|p{1cm}|p{1.6cm}|p{1.6cm}|p{1.9cm}}
\hline
states & LB fit (a.u.) & (a) (a.u.) & (b) (a.u.) & (c) (a.u.)\\
\hline	
4(1) 			& -8238 	& -7484.5		& -9351 	& -9437 	\\
1(2) 			& -7236 	&	-7484.5		& -9351 	& -9428 	\\
3(1) 			& -14509	& -14909		& -17702 	& -18261  \\
3(0$^-$) 	& -14509 	& -14909		& -17702 	& -18276  \\
3(0$^+$) 	& -12498 	& -14909		& -17702 	& -18252  \\
\hline
\end{tabular}
\label{table2}
\end{table} 

The uncertainties of our results for the $C_6$ coefficients can be discussed by estimating the contribution of the next term in the multipolar expansion of the long-range interaction energy, i.e $C_8/R^8$. At the internuclear distance of 30 $a_0$ already reached by several of the observed vibrational levels, and using $C_8$ values from Ref.\cite{Marinescu1999}, this term amounts for a noticeable fraction of about 11$\%$ of the van der Waals energy. However, a fitting using an improved LB formula~\cite{Comparat2004} including both $C_6/R^6$ and $C_8/R^8$ contributions does not give satisfactory results as there are too many free parameters compared to the number of observed lines. We believe that scanning for deeper bound levels will not help to improve the results, as we are already close to the internuclear distance where exchange potential and higher order terms in the multipolar expansion should play a role. Thus we will have uncertainties on our $C_6$ values because we are at the limit of what we can do with the LB formula. 

We also note that the LB formula should be applied to molecules in zero magnetic field only, while our spectra are taken at 340 G. The Zeeman and hyperfine interactions result in complicated structures corresponding to successive series converging in principle toward different hyperfine Zeeman dissociation limits, such that application of the LB formula, strictly speaking, is impossible. In this work, we defined a center of gravity for each structured line assuming equal weighting of all observed substructures, and the detuning (binding energy) corresponding to this center of gravity is determined with respect to the single dissociation limit as defined in Sec.\ref{experiment}. Although both the Zeeman shift and the overall substructure width are much smaller than the vibrational level spaces, the $C_6$ values extracted could still be slightly distorted.

\section{Conclusion}
 \label{conclusion}
A sample of ultracold Feshbach molecules is a good starting point for investigating excited-state molecular potentials with very high resolution, especially for long-range states hard to access with thermal samples. In this work we have investigated the near dissociation levels of the $^{23}$Na$^{87}$Rb molecule correlated with the Na($3^{2}S_{1/2}$)+Rb($5^{2}P_{3/2}$) asymptote and observed the 5 long-range states with attractive potentials. We have extracted the $C_6$ coefficients with the LB model and found reasonable agreements with previous theoretical calculations.

However, the current study is already covering internuclear distances violating the long range model. Thus the experimental $C_6$ parameters are a kind of ``effective'' parameters which already include significant contributions from short-range potentials. Similar difficulties were already met on the LiK long-range state analysis~\cite{Ridinger2011}. This situation is probably quite true for such molecules with light reduced masses (and thus a smaller density of states). To better match the experimental values, the asymptotic models for calculating the van der Waals coefficients will need some refinements. In particular, such models predict identical $C_6$ values for molecular states~\cite{Marinescu1999,Frecon1998}, which is not what we found experimentally.

In future works, extension of this spectroscopy to much deeper levels matching the range already addressed by conventional spectroscopy should be feasible. Combining these works together will lead to a full understanding of the relevant excited states and help us identify the suitable intermediate levels for Raman population transfer to the ground state.

\section{Acknowledgments}
We thank Xin Ye and Bo Lu for laboratory assistance, and W. C. Stwalley for stimulating discussions. This work is supported by the COPOMOL project which is jointly funded by Hong Kong RGC (grant no. A-CUHK403/13) and French ANR (grant no. ANR-13-IS04-0004-01). The Hong Kong team is also supported by the RGC General Research Fund (grant no. CUHK404712) and the National Basic Research Program of China (grant No. 2014CB921402).


\begin{thebibliography}{35}%
\makeatletter
\providecommand \@ifxundefined [1]{%
 \@ifx{#1\undefined}
}%
\providecommand \@ifnum [1]{%
 \ifnum #1\expandafter \@firstoftwo
 \else \expandafter \@secondoftwo
 \fi
}%
\providecommand \@ifx [1]{%
 \ifx #1\expandafter \@firstoftwo
 \else \expandafter \@secondoftwo
 \fi
}%
\providecommand \natexlab [1]{#1}%
\providecommand \enquote  [1]{``#1''}%
\providecommand \bibnamefont  [1]{#1}%
\providecommand \bibfnamefont [1]{#1}%
\providecommand \citenamefont [1]{#1}%
\providecommand \href@noop [0]{\@secondoftwo}%
\providecommand \href [0]{\begingroup \@sanitize@url \@href}%
\providecommand \@href[1]{\@@startlink{#1}\@@href}%
\providecommand \@@href[1]{\endgroup#1\@@endlink}%
\providecommand \@sanitize@url [0]{\catcode `\\12\catcode `\$12\catcode
  `\&12\catcode `\#12\catcode `\^12\catcode `\_12\catcode `\%12\relax}%
\providecommand \@@startlink[1]{}%
\providecommand \@@endlink[0]{}%
\providecommand \url  [0]{\begingroup\@sanitize@url \@url }%
\providecommand \@url [1]{\endgroup\@href {#1}{\urlprefix }}%
\providecommand \urlprefix  [0]{URL }%
\providecommand \Eprint [0]{\href }%
\providecommand \doibase [0]{http://dx.doi.org/}%
\providecommand \selectlanguage [0]{\@gobble}%
\providecommand \bibinfo  [0]{\@secondoftwo}%
\providecommand \bibfield  [0]{\@secondoftwo}%
\providecommand \translation [1]{[#1]}%
\providecommand \BibitemOpen [0]{}%
\providecommand \bibitemStop [0]{}%
\providecommand \bibitemNoStop [0]{.\EOS\space}%
\providecommand \EOS [0]{\spacefactor3000\relax}%
\providecommand \BibitemShut  [1]{\csname bibitem#1\endcsname}%
\let\auto@bib@innerbib\@empty
\bibitem [{\citenamefont {Doyle}\ \emph {et~al.}(2004)\citenamefont {Doyle},
  \citenamefont {Friedrich}, \citenamefont {Krems},\ and\ \citenamefont
  {Masnou-Seeuws}}]{EPJD2004}%
  \BibitemOpen
  \bibinfo {editor} {\bibfnamefont {J.}~\bibnamefont {Doyle}}, \bibinfo
  {editor} {\bibfnamefont {B.}~\bibnamefont {Friedrich}}, \bibinfo {editor}
  {\bibfnamefont {R.}~\bibnamefont {Krems}}, \ and\ \bibinfo {editor}
  {\bibfnamefont {F.}~\bibnamefont {Masnou-Seeuws}},\ eds.,\ \href@noop {}
  {\emph {\bibinfo {title} {Special Issue: ``Ultracold Polar Molecules:
  Formation and Collisions''}}},\ Vol.~\bibinfo {volume} {31}\ (\bibinfo
  {publisher} {Eur. Phys. J. D},\ \bibinfo {year} {2004})\BibitemShut {NoStop}%
\bibitem [{\citenamefont {Carr}\ and\ \citenamefont {Ye}(2009)}]{Carr2009}%
  \BibitemOpen
  \bibinfo {editor} {\bibfnamefont {L.~D.}\ \bibnamefont {Carr}}\ and\ \bibinfo
  {editor} {\bibfnamefont {J.}~\bibnamefont {Ye}},\ eds.,\ \href@noop {} {\emph
  {\bibinfo {title} {Special Issue: ``Focus on Cold and Ultracold
  Molecules''}}},\ Vol.~\bibinfo {volume} {11}\ (\bibinfo  {publisher} {New J.
  Phys.},\ \bibinfo {year} {2009})\BibitemShut {NoStop}%
\bibitem [{\citenamefont {Shuman}\ \emph {et~al.}(2010)\citenamefont {Shuman},
  \citenamefont {Barry},\ and\ \citenamefont {DeMille.}}]{Shuman2010}%
  \BibitemOpen
  \bibfield  {author} {\bibinfo {author} {\bibfnamefont {E.~S.}\ \bibnamefont
  {Shuman}}, \bibinfo {author} {\bibfnamefont {J.}~\bibnamefont {Barry}}, \
  and\ \bibinfo {author} {\bibfnamefont {D.}~\bibnamefont {DeMille}},\
  }\href@noop {} {\bibfield  {journal} {\bibinfo  {journal} {Nature}\ }\textbf
  {\bibinfo {volume} {467}},\ \bibinfo {pages} {820} (\bibinfo {year}
  {2010})}\BibitemShut {NoStop}%
\bibitem [{\citenamefont {Hummon}\ \emph {et~al.}(2013)\citenamefont {Hummon},
  \citenamefont {Yeo}, \citenamefont {Stuhl}, \citenamefont {Collopy},
  \citenamefont {Xia},\ and\ \citenamefont {Ye}}]{Hummon2013}%
  \BibitemOpen
  \bibfield  {author} {\bibinfo {author} {\bibfnamefont {M.~T.}\ \bibnamefont
  {Hummon}}, \bibinfo {author} {\bibfnamefont {M.}~\bibnamefont {Yeo}},
  \bibinfo {author} {\bibfnamefont {B.~K.}\ \bibnamefont {Stuhl}}, \bibinfo
  {author} {\bibfnamefont {A.~L.}\ \bibnamefont {Collopy}}, \bibinfo {author}
  {\bibfnamefont {Y.}~\bibnamefont {Xia}}, \ and\ \bibinfo {author}
  {\bibfnamefont {J.}~\bibnamefont {Ye}},\ }\href {\doibase
  10.1103/PhysRevLett.110.143001} {\bibfield  {journal} {\bibinfo  {journal}
  {Phys. Rev. Lett.}\ }\textbf {\bibinfo {volume} {110}},\ \bibinfo {pages}
  {143001} (\bibinfo {year} {2013})}\BibitemShut {NoStop}%
\bibitem [{\citenamefont {Barry}\ \emph {et~al.}(2014)\citenamefont {Barry},
  \citenamefont {McCarron}, \citenamefont {Norrgard}, \citenamefont
  {Steinecker},\ and\ \citenamefont {DeMille}}]{Barry2014}%
  \BibitemOpen
  \bibfield  {author} {\bibinfo {author} {\bibfnamefont {J.~F.}\ \bibnamefont
  {Barry}}, \bibinfo {author} {\bibfnamefont {D.~J.}\ \bibnamefont {McCarron}},
  \bibinfo {author} {\bibfnamefont {E.~B.}\ \bibnamefont {Norrgard}}, \bibinfo
  {author} {\bibfnamefont {M.~H.}\ \bibnamefont {Steinecker}}, \ and\ \bibinfo
  {author} {\bibfnamefont {D.}~\bibnamefont {DeMille}},\ }\href@noop {}
  {\bibfield  {journal} {\bibinfo  {journal} {Nature}\ }\textbf {\bibinfo
  {volume} {512}},\ \bibinfo {pages} {286} (\bibinfo {year}
  {2014})}\BibitemShut {NoStop}%
\bibitem [{\citenamefont {K\"{o}hler}\ \emph {et~al.}(2006)\citenamefont
  {K\"{o}hler}, \citenamefont {G\'{o}ral},\ and\ \citenamefont
  {Julienne}}]{Kohler06}%
  \BibitemOpen
  \bibfield  {author} {\bibinfo {author} {\bibfnamefont {T.}~\bibnamefont
  {K\"{o}hler}}, \bibinfo {author} {\bibfnamefont {K.}~\bibnamefont
  {G\'{o}ral}}, \ and\ \bibinfo {author} {\bibfnamefont {P.~S.}\ \bibnamefont
  {Julienne}},\ }\href@noop {} {\bibfield  {journal} {\bibinfo  {journal} {Rev.
  Mod. Phys.}\ }\textbf {\bibinfo {volume} {78}},\ \bibinfo {eid} {1311}
  (\bibinfo {year} {2006})}\BibitemShut {NoStop}%
\bibitem [{\citenamefont {Winkler}\ \emph {et~al.}(2007)\citenamefont
  {Winkler}, \citenamefont {Lang}, \citenamefont {Thalhammer}, \citenamefont
  {v.~d. Straten}, \citenamefont {Grimm},\ and\ \citenamefont
  {Denschlag}}]{Winkler2006}%
  \BibitemOpen
  \bibfield  {author} {\bibinfo {author} {\bibfnamefont {K.}~\bibnamefont
  {Winkler}}, \bibinfo {author} {\bibfnamefont {F.}~\bibnamefont {Lang}},
  \bibinfo {author} {\bibfnamefont {G.}~\bibnamefont {Thalhammer}}, \bibinfo
  {author} {\bibfnamefont {P.}~\bibnamefont {v.~d. Straten}}, \bibinfo {author}
  {\bibfnamefont {R.}~\bibnamefont {Grimm}}, \ and\ \bibinfo {author}
  {\bibfnamefont {J.~H.}\ \bibnamefont {Denschlag}},\ }\href
  {http://link.aps.org/abstract/PRL/v98/e043201} {\bibfield  {journal}
  {\bibinfo  {journal} {Phys. Rev. Lett.}\ }\textbf {\bibinfo {volume} {98}},\
  \bibinfo {eid} {043201} (\bibinfo {year} {2007})}\BibitemShut {NoStop}%
\bibitem [{\citenamefont {Danzl}\ \emph {et~al.}(2008)\citenamefont {Danzl},
  \citenamefont {Haller}, \citenamefont {Gustavsson}, \citenamefont {Mark},
  \citenamefont {Hart}, \citenamefont {Bouloufa}, \citenamefont {Dulieu},
  \citenamefont {Ritsch},\ and\ \citenamefont {N\"agerl}}]{Danzl2008}%
  \BibitemOpen
  \bibfield  {author} {\bibinfo {author} {\bibfnamefont {J.~G.}\ \bibnamefont
  {Danzl}}, \bibinfo {author} {\bibfnamefont {E.}~\bibnamefont {Haller}},
  \bibinfo {author} {\bibfnamefont {M.}~\bibnamefont {Gustavsson}}, \bibinfo
  {author} {\bibfnamefont {M.~J.}\ \bibnamefont {Mark}}, \bibinfo {author}
  {\bibfnamefont {R.}~\bibnamefont {Hart}}, \bibinfo {author} {\bibfnamefont
  {N.}~\bibnamefont {Bouloufa}}, \bibinfo {author} {\bibfnamefont
  {O.}~\bibnamefont {Dulieu}}, \bibinfo {author} {\bibfnamefont
  {H.}~\bibnamefont {Ritsch}}, \ and\ \bibinfo {author} {\bibfnamefont {H.-C.}\
  \bibnamefont {N\"agerl}},\ }\href {\doibase 10.1126/science.1159909}
  {\bibfield  {journal} {\bibinfo  {journal} {Science}\ }\textbf {\bibinfo
  {volume} {321}},\ \bibinfo {pages} {1062} (\bibinfo {year}
  {2008})}\BibitemShut {NoStop}%
\bibitem [{\citenamefont {Ospelkaus}\ \emph {et~al.}(2008)\citenamefont
  {Ospelkaus}, \citenamefont {Peӥr}, \citenamefont {Ni}, \citenamefont
  {Zirbel}, \citenamefont {Neyenhuis}, \citenamefont {Kotochigova},
  \citenamefont {Julienne}, \citenamefont {Ye},\ and\ \citenamefont
  {Jin}}]{Ospelkaus2008}%
  \BibitemOpen
  \bibfield  {author} {\bibinfo {author} {\bibfnamefont {S.}~\bibnamefont
  {Ospelkaus}}, \bibinfo {author} {\bibfnamefont {A.}~\bibnamefont {Peӥr}},
  \bibinfo {author} {\bibfnamefont {K.~K.}\ \bibnamefont {Ni}}, \bibinfo
  {author} {\bibfnamefont {J.~J.}\ \bibnamefont {Zirbel}}, \bibinfo {author}
  {\bibfnamefont {B.}~\bibnamefont {Neyenhuis}}, \bibinfo {author}
  {\bibfnamefont {S.}~\bibnamefont {Kotochigova}}, \bibinfo {author}
  {\bibfnamefont {P.~S.}\ \bibnamefont {Julienne}}, \bibinfo {author}
  {\bibfnamefont {J.}~\bibnamefont {Ye}}, \ and\ \bibinfo {author}
  {\bibfnamefont {D.~S.}\ \bibnamefont {Jin}},\ }\href@noop {} {\bibfield
  {journal} {\bibinfo  {journal} {Nature Phys.}\ }\textbf {\bibinfo {volume}
  {4}},\ \bibinfo {pages} {622} (\bibinfo {year} {2008})}\BibitemShut {NoStop}%
\bibitem [{\citenamefont {Ni}\ \emph {et~al.}(2008)\citenamefont {Ni},
  \citenamefont {Ospelkaus}, \citenamefont {de~Miranda}, \citenamefont {Pe'er},
  \citenamefont {Neyenhuis}, \citenamefont {Zirbel}, \citenamefont
  {Kotochigova}, \citenamefont {Julienne}, \citenamefont {Jin},\ and\
  \citenamefont {Ye}}]{Ni2008}%
  \BibitemOpen
  \bibfield  {author} {\bibinfo {author} {\bibfnamefont {K.-K.}\ \bibnamefont
  {Ni}}, \bibinfo {author} {\bibfnamefont {S.}~\bibnamefont {Ospelkaus}},
  \bibinfo {author} {\bibfnamefont {M.~H.~G.}\ \bibnamefont {de~Miranda}},
  \bibinfo {author} {\bibfnamefont {A.}~\bibnamefont {Pe'er}}, \bibinfo
  {author} {\bibfnamefont {B.}~\bibnamefont {Neyenhuis}}, \bibinfo {author}
  {\bibfnamefont {J.~J.}\ \bibnamefont {Zirbel}}, \bibinfo {author}
  {\bibfnamefont {S.}~\bibnamefont {Kotochigova}}, \bibinfo {author}
  {\bibfnamefont {P.~S.}\ \bibnamefont {Julienne}}, \bibinfo {author}
  {\bibfnamefont {D.~S.}\ \bibnamefont {Jin}}, \ and\ \bibinfo {author}
  {\bibfnamefont {J.}~\bibnamefont {Ye}},\ }\href {\doibase
  10.1126/science.1163861} {\bibfield  {journal} {\bibinfo  {journal}
  {Science}\ }\textbf {\bibinfo {volume} {322}},\ \bibinfo {pages} {231}
  (\bibinfo {year} {2008})}\BibitemShut {NoStop}%
\bibitem [{\citenamefont {Takekoshi}\ \emph {et~al.}(2014)\citenamefont
  {Takekoshi}, \citenamefont {Reichs\"ollner}, \citenamefont {Schindewolf},
  \citenamefont {Hutson}, \citenamefont {Le~Sueur}, \citenamefont {Dulieu},
  \citenamefont {Ferlaino}, \citenamefont {Grimm},\ and\ \citenamefont
  {N\"agerl}}]{Takekoshi2014}%
  \BibitemOpen
  \bibfield  {author} {\bibinfo {author} {\bibfnamefont {T.}~\bibnamefont
  {Takekoshi}}, \bibinfo {author} {\bibfnamefont {L.}~\bibnamefont
  {Reichs\"ollner}}, \bibinfo {author} {\bibfnamefont {A.}~\bibnamefont
  {Schindewolf}}, \bibinfo {author} {\bibfnamefont {J.~M.}\ \bibnamefont
  {Hutson}}, \bibinfo {author} {\bibfnamefont {C.~R.}\ \bibnamefont
  {Le~Sueur}}, \bibinfo {author} {\bibfnamefont {O.}~\bibnamefont {Dulieu}},
  \bibinfo {author} {\bibfnamefont {F.}~\bibnamefont {Ferlaino}}, \bibinfo
  {author} {\bibfnamefont {R.}~\bibnamefont {Grimm}}, \ and\ \bibinfo {author}
  {\bibfnamefont {H.-C.}\ \bibnamefont {N\"agerl}},\ }\href {\doibase
  10.1103/PhysRevLett.113.205301} {\bibfield  {journal} {\bibinfo  {journal}
  {Phys. Rev. Lett.}\ }\textbf {\bibinfo {volume} {113}},\ \bibinfo {pages}
  {205301} (\bibinfo {year} {2014})}\BibitemShut {NoStop}%
\bibitem [{\citenamefont {Molony}\ \emph {et~al.}(2014)\citenamefont {Molony},
  \citenamefont {Gregory}, \citenamefont {Ji}, \citenamefont {Lu},
  \citenamefont {K\"oppinger}, \citenamefont {Le~Sueur}, \citenamefont
  {Blackley}, \citenamefont {Hutson},\ and\ \citenamefont
  {Cornish}}]{Molony2014}%
  \BibitemOpen
  \bibfield  {author} {\bibinfo {author} {\bibfnamefont {P.~K.}\ \bibnamefont
  {Molony}}, \bibinfo {author} {\bibfnamefont {P.~D.}\ \bibnamefont {Gregory}},
  \bibinfo {author} {\bibfnamefont {Z.}~\bibnamefont {Ji}}, \bibinfo {author}
  {\bibfnamefont {B.}~\bibnamefont {Lu}}, \bibinfo {author} {\bibfnamefont
  {M.~P.}\ \bibnamefont {K\"oppinger}}, \bibinfo {author} {\bibfnamefont
  {C.~R.}\ \bibnamefont {Le~Sueur}}, \bibinfo {author} {\bibfnamefont {C.~L.}\
  \bibnamefont {Blackley}}, \bibinfo {author} {\bibfnamefont {J.~M.}\
  \bibnamefont {Hutson}}, \ and\ \bibinfo {author} {\bibfnamefont {S.~L.}\
  \bibnamefont {Cornish}},\ }\href {\doibase 10.1103/PhysRevLett.113.255301}
  {\bibfield  {journal} {\bibinfo  {journal} {Phys. Rev. Lett.}\ }\textbf
  {\bibinfo {volume} {113}},\ \bibinfo {pages} {255301} (\bibinfo {year}
  {2014})}\BibitemShut {NoStop}%
\bibitem [{\citenamefont {Park}\ \emph {et~al.}(2015)\citenamefont {Park},
  \citenamefont {Will},\ and\ \citenamefont {Zwierlein}}]{Park2015}%
  \BibitemOpen
  \bibfield  {author} {\bibinfo {author} {\bibfnamefont {J.~W.}\ \bibnamefont
  {Park}}, \bibinfo {author} {\bibfnamefont {S.~A.}\ \bibnamefont {Will}}, \
  and\ \bibinfo {author} {\bibfnamefont {M.~W.}\ \bibnamefont {Zwierlein}},\
  }\href {\doibase 10.1103/PhysRevLett.114.205302} {\bibfield  {journal}
  {\bibinfo  {journal} {Phys. Rev. Lett.}\ }\textbf {\bibinfo {volume} {114}},\
  \bibinfo {pages} {205302} (\bibinfo {year} {2015})}\BibitemShut {NoStop}%
\bibitem [{\citenamefont {Ospelkaus}\ \emph {et~al.}(2010)\citenamefont
  {Ospelkaus}, \citenamefont {Ni}, \citenamefont {Wang}, \citenamefont
  {de~Miranda}, \citenamefont {Neyenhuis}, \citenamefont {Qu{\'e}m{\'e}ner},
  \citenamefont {Julienne}, \citenamefont {Bohn}, \citenamefont {Jin},\ and\
  \citenamefont {Ye}}]{Ospelkaus10}%
  \BibitemOpen
  \bibfield  {author} {\bibinfo {author} {\bibfnamefont {S.}~\bibnamefont
  {Ospelkaus}}, \bibinfo {author} {\bibfnamefont {K.-K.}\ \bibnamefont {Ni}},
  \bibinfo {author} {\bibfnamefont {D.}~\bibnamefont {Wang}}, \bibinfo {author}
  {\bibfnamefont {M.~H.~G.}\ \bibnamefont {de~Miranda}}, \bibinfo {author}
  {\bibfnamefont {B.}~\bibnamefont {Neyenhuis}}, \bibinfo {author}
  {\bibfnamefont {G.}~\bibnamefont {Qu{\'e}m{\'e}ner}}, \bibinfo {author}
  {\bibfnamefont {P.~S.}\ \bibnamefont {Julienne}}, \bibinfo {author}
  {\bibfnamefont {J.~L.}\ \bibnamefont {Bohn}}, \bibinfo {author}
  {\bibfnamefont {D.~S.}\ \bibnamefont {Jin}}, \ and\ \bibinfo {author}
  {\bibfnamefont {J.}~\bibnamefont {Ye}},\ }\href@noop {} {\bibfield  {journal}
  {\bibinfo  {journal} {Science}\ }\textbf {\bibinfo {volume} {327}},\ \bibinfo
  {pages} {853} (\bibinfo {year} {2010})}\BibitemShut {NoStop}%
\bibitem [{\citenamefont {Ni}\ \emph {et~al.}(2010)\citenamefont {Ni},
  \citenamefont {Ospelkaus}, \citenamefont {Wang}, \citenamefont
  {Qu{\'e}m{\'e}ner}, \citenamefont {Neyenhuis}, \citenamefont {de~Miranda},
  \citenamefont {Bohn}, \citenamefont {Ye},\ and\ \citenamefont {Jin}}]{Ni10}%
  \BibitemOpen
  \bibfield  {author} {\bibinfo {author} {\bibfnamefont {K.-K.}\ \bibnamefont
  {Ni}}, \bibinfo {author} {\bibfnamefont {S.}~\bibnamefont {Ospelkaus}},
  \bibinfo {author} {\bibfnamefont {D.}~\bibnamefont {Wang}}, \bibinfo {author}
  {\bibfnamefont {G.}~\bibnamefont {Qu{\'e}m{\'e}ner}}, \bibinfo {author}
  {\bibfnamefont {B.}~\bibnamefont {Neyenhuis}}, \bibinfo {author}
  {\bibfnamefont {M.~H.~G.}\ \bibnamefont {de~Miranda}}, \bibinfo {author}
  {\bibfnamefont {J.~L.}\ \bibnamefont {Bohn}}, \bibinfo {author}
  {\bibfnamefont {J.}~\bibnamefont {Ye}}, \ and\ \bibinfo {author}
  {\bibfnamefont {D.~S.}\ \bibnamefont {Jin}},\ }\href@noop {} {\bibfield
  {journal} {\bibinfo  {journal} {Nature}\ }\textbf {\bibinfo {volume} {464}},\
  \bibinfo {pages} {1324} (\bibinfo {year} {2010})}\BibitemShut {NoStop}%
\bibitem [{\citenamefont {\ifmmode~\dot{Z}\else \.{Z}\fi{}uchowski}\ and\
  \citenamefont {Hutson}(2010)}]{Zuchowski2010}%
  \BibitemOpen
  \bibfield  {author} {\bibinfo {author} {\bibfnamefont {P.~S.}\ \bibnamefont
  {\ifmmode~\dot{Z}\else \.{Z}\fi{}uchowski}}\ and\ \bibinfo {author}
  {\bibfnamefont {J.~M.}\ \bibnamefont {Hutson}},\ }\href {\doibase
  10.1103/PhysRevA.81.060703} {\bibfield  {journal} {\bibinfo  {journal} {Phys.
  Rev. A}\ }\textbf {\bibinfo {volume} {81}},\ \bibinfo {pages} {060703}
  (\bibinfo {year} {2010})}\BibitemShut {NoStop}%
\bibitem [{\citenamefont {Mayle}\ \emph {et~al.}(2013)\citenamefont {Mayle},
  \citenamefont {Qu\'em\'ener}, \citenamefont {Ruzic},\ and\ \citenamefont
  {Bohn}}]{Mayle2013}%
  \BibitemOpen
  \bibfield  {author} {\bibinfo {author} {\bibfnamefont {M.}~\bibnamefont
  {Mayle}}, \bibinfo {author} {\bibfnamefont {G.}~\bibnamefont {Qu\'em\'ener}},
  \bibinfo {author} {\bibfnamefont {B.~P.}\ \bibnamefont {Ruzic}}, \ and\
  \bibinfo {author} {\bibfnamefont {J.~L.}\ \bibnamefont {Bohn}},\ }\href
  {\doibase 10.1103/PhysRevA.87.012709} {\bibfield  {journal} {\bibinfo
  {journal} {Phys. Rev. A}\ }\textbf {\bibinfo {volume} {87}},\ \bibinfo
  {pages} {012709} (\bibinfo {year} {2013})}\BibitemShut {NoStop}%
\bibitem [{\citenamefont {Aymar}\ and\ \citenamefont
  {Dulieu}(2005)}]{Aymar2005}%
  \BibitemOpen
  \bibfield  {author} {\bibinfo {author} {\bibfnamefont {M.}~\bibnamefont
  {Aymar}}\ and\ \bibinfo {author} {\bibfnamefont {O.}~\bibnamefont {Dulieu}},\
  }\href {http://link.aip.org/link/?JCP/122/204302/1} {\bibfield  {journal}
  {\bibinfo  {journal} {J. Chem. Phys.}\ }\textbf {\bibinfo {volume} {122}},\
  \bibinfo {eid} {204302} (\bibinfo {year} {2005})}\BibitemShut {NoStop}%
\bibitem [{\citenamefont {Xiong}\ \emph {et~al.}()\citenamefont {Xiong},
  \citenamefont {Wang}, \citenamefont {Li}, \citenamefont {Lam},\ and\
  \citenamefont {Wang}}]{Xiong2013}%
  \BibitemOpen
  \bibfield  {author} {\bibinfo {author} {\bibfnamefont {D.}~\bibnamefont
  {Xiong}}, \bibinfo {author} {\bibfnamefont {F.}~\bibnamefont {Wang}},
  \bibinfo {author} {\bibfnamefont {X.}~\bibnamefont {Li}}, \bibinfo {author}
  {\bibfnamefont {T.-F.}\ \bibnamefont {Lam}}, \ and\ \bibinfo {author}
  {\bibfnamefont {D.}~\bibnamefont {Wang}},\ }\href@noop {} {\bibinfo
  {journal} {arXiv:1303.0333}\ }\BibitemShut {NoStop}%
\bibitem [{\citenamefont {Wang}\ \emph {et~al.}(2013)\citenamefont {Wang},
  \citenamefont {Xiong}, \citenamefont {Li}, \citenamefont {Wang},\ and\
  \citenamefont {Tiemann}}]{Wangfudong2013}%
  \BibitemOpen
\bibfield  {journal} {  }\bibfield  {author} {\bibinfo {author} {\bibfnamefont
  {F.}~\bibnamefont {Wang}}, \bibinfo {author} {\bibfnamefont {D.}~\bibnamefont
  {Xiong}}, \bibinfo {author} {\bibfnamefont {X.}~\bibnamefont {Li}}, \bibinfo
  {author} {\bibfnamefont {D.}~\bibnamefont {Wang}}, \ and\ \bibinfo {author}
  {\bibfnamefont {E.}~\bibnamefont {Tiemann}},\ }\href {\doibase
  10.1103/PhysRevA.87.050702} {\bibfield  {journal} {\bibinfo  {journal} {Phys.
  Rev. A(R)}\ }\textbf {\bibinfo {volume} {87}},\ \bibinfo {pages} {050702}
  (\bibinfo {year} {2013})}\BibitemShut {NoStop}%
\bibitem [{\citenamefont {Wang}\ \emph {et~al.}(2015)\citenamefont {Wang},
  \citenamefont {He}, \citenamefont {Li}, \citenamefont {Zhu}, \citenamefont
  {Chen},\ and\ \citenamefont {Wang}}]{Wangfudong2015}%
  \BibitemOpen
  \bibfield  {author} {\bibinfo {author} {\bibfnamefont {F.}~\bibnamefont
  {Wang}}, \bibinfo {author} {\bibfnamefont {X.}~\bibnamefont {He}}, \bibinfo
  {author} {\bibfnamefont {X.}~\bibnamefont {Li}}, \bibinfo {author}
  {\bibfnamefont {B.}~\bibnamefont {Zhu}}, \bibinfo {author} {\bibfnamefont
  {J.}~\bibnamefont {Chen}}, \ and\ \bibinfo {author} {\bibfnamefont
  {D.}~\bibnamefont {Wang}},\ }\href
  {http://stacks.iop.org/1367-2630/17/i=3/a=035003} {\bibfield  {journal}
  {\bibinfo  {journal} {New J. Phys.}\ }\textbf {\bibinfo {volume} {17}},\
  \bibinfo {pages} {035003} (\bibinfo {year} {2015})}\BibitemShut {NoStop}%
\bibitem [{\citenamefont {Pashov}\ \emph {et~al.}(2005)\citenamefont {Pashov},
  \citenamefont {Docenko}, \citenamefont {Tamanis}, \citenamefont {Ferber},
  \citenamefont {Kn\"ockel},\ and\ \citenamefont {Tiemann}}]{Pashov2005}%
  \BibitemOpen
  \bibfield  {author} {\bibinfo {author} {\bibfnamefont {A.}~\bibnamefont
  {Pashov}}, \bibinfo {author} {\bibfnamefont {O.}~\bibnamefont {Docenko}},
  \bibinfo {author} {\bibfnamefont {M.}~\bibnamefont {Tamanis}}, \bibinfo
  {author} {\bibfnamefont {R.}~\bibnamefont {Ferber}}, \bibinfo {author}
  {\bibfnamefont {H.}~\bibnamefont {Kn\"ockel}}, \ and\ \bibinfo {author}
  {\bibfnamefont {E.}~\bibnamefont {Tiemann}},\ }\href {\doibase
  10.1103/PhysRevA.72.062505} {\bibfield  {journal} {\bibinfo  {journal} {Phys.
  Rev. A}\ }\textbf {\bibinfo {volume} {72}},\ \bibinfo {pages} {062505}
  (\bibinfo {year} {2005})}\BibitemShut {NoStop}%
\bibitem [{\citenamefont {Wang}\ \emph {et~al.}(1991)\citenamefont {Wang},
  \citenamefont {Kajitani}, \citenamefont {Kasahara}, \citenamefont {Baba},
  \citenamefont {Ishikawa},\ and\ \citenamefont {Kato}}]{Wangyouchang1991}%
  \BibitemOpen
  \bibfield  {author} {\bibinfo {author} {\bibfnamefont {Y.-C.}\ \bibnamefont
  {Wang}}, \bibinfo {author} {\bibfnamefont {M.}~\bibnamefont {Kajitani}},
  \bibinfo {author} {\bibfnamefont {S.}~\bibnamefont {Kasahara}}, \bibinfo
  {author} {\bibfnamefont {M.}~\bibnamefont {Baba}}, \bibinfo {author}
  {\bibfnamefont {K.}~\bibnamefont {Ishikawa}}, \ and\ \bibinfo {author}
  {\bibfnamefont {H.}~\bibnamefont {Kato}},\ }\href {\doibase
  http://dx.doi.org/10.1063/1.461569} {\bibfield  {journal} {\bibinfo
  {journal} {J. Chem. Phys.}\ }\textbf {\bibinfo {volume} {95}},\ \bibinfo
  {pages} {6229} (\bibinfo {year} {1991})}\BibitemShut {NoStop}%
\bibitem [{\citenamefont {Pashov}\ \emph {et~al.}(2006)\citenamefont {Pashov},
  \citenamefont {Jastrzebski}, \citenamefont {Kortyka},\ and\ \citenamefont
  {Kowalczyk}}]{Pashov2006}%
  \BibitemOpen
  \bibfield  {author} {\bibinfo {author} {\bibfnamefont {A.}~\bibnamefont
  {Pashov}}, \bibinfo {author} {\bibfnamefont {W.}~\bibnamefont {Jastrzebski}},
  \bibinfo {author} {\bibfnamefont {P.}~\bibnamefont {Kortyka}}, \ and\
  \bibinfo {author} {\bibfnamefont {P.}~\bibnamefont {Kowalczyk}},\ }\href
  {\doibase http://dx.doi.org/10.1063/1.2198199} {\bibfield  {journal}
  {\bibinfo  {journal} {J. Chem. Phys.}\ }\textbf {\bibinfo {volume} {124}},\
  \bibinfo {eid} {204308} (\bibinfo {year} {2006})}\BibitemShut {NoStop}%
\bibitem [{\citenamefont {Docenko}\ \emph {et~al.}(2007)\citenamefont
  {Docenko}, \citenamefont {Tamanis}, \citenamefont {Ferber}, \citenamefont
  {Pazyuk}, \citenamefont {Zaitsevskii}, \citenamefont {Stolyarov},
  \citenamefont {Pashov}, \citenamefont {Kn\"ockel},\ and\ \citenamefont
  {Tiemann}}]{Docenko2007}%
  \BibitemOpen
  \bibfield  {author} {\bibinfo {author} {\bibfnamefont {O.}~\bibnamefont
  {Docenko}}, \bibinfo {author} {\bibfnamefont {M.}~\bibnamefont {Tamanis}},
  \bibinfo {author} {\bibfnamefont {R.}~\bibnamefont {Ferber}}, \bibinfo
  {author} {\bibfnamefont {E.~A.}\ \bibnamefont {Pazyuk}}, \bibinfo {author}
  {\bibfnamefont {A.}~\bibnamefont {Zaitsevskii}}, \bibinfo {author}
  {\bibfnamefont {A.~V.}\ \bibnamefont {Stolyarov}}, \bibinfo {author}
  {\bibfnamefont {A.}~\bibnamefont {Pashov}}, \bibinfo {author} {\bibfnamefont
  {H.}~\bibnamefont {Kn\"ockel}}, \ and\ \bibinfo {author} {\bibfnamefont
  {E.}~\bibnamefont {Tiemann}},\ }\href {\doibase 10.1103/PhysRevA.75.042503}
  {\bibfield  {journal} {\bibinfo  {journal} {Phys. Rev. A}\ }\textbf {\bibinfo
  {volume} {75}},\ \bibinfo {pages} {042503} (\bibinfo {year}
  {2007})}\BibitemShut {NoStop}%
\bibitem [{\citenamefont {Korek}\ and\ \citenamefont
  {Fawwaz}(2009)}]{Korek2009}%
  \BibitemOpen
  \bibfield  {author} {\bibinfo {author} {\bibfnamefont {M.}~\bibnamefont
  {Korek}}\ and\ \bibinfo {author} {\bibfnamefont {O.}~\bibnamefont {Fawwaz}},\
  }\href {\doibase 10.1002/qua.21904} {\bibfield  {journal} {\bibinfo
  {journal} {Int. J. Quant. Chem.}\ }\textbf {\bibinfo {volume} {109}},\
  \bibinfo {pages} {938} (\bibinfo {year} {2009})}\BibitemShut {NoStop}%
\bibitem [{\citenamefont {Borsalino}\ \emph {et~al.}(2014)\citenamefont
  {Borsalino}, \citenamefont {Londo\~no Flor\`ez}, \citenamefont {Vexiau},
  \citenamefont {Dulieu}, \citenamefont {Bouloufa-Maafa},\ and\ \citenamefont
  {Luc-Koenig}}]{Borsalino2014}%
  \BibitemOpen
  \bibfield  {author} {\bibinfo {author} {\bibfnamefont {D.}~\bibnamefont
  {Borsalino}}, \bibinfo {author} {\bibfnamefont {B.}~\bibnamefont {Londo\~no
  Flor\`ez}}, \bibinfo {author} {\bibfnamefont {R.}~\bibnamefont {Vexiau}},
  \bibinfo {author} {\bibfnamefont {O.}~\bibnamefont {Dulieu}}, \bibinfo
  {author} {\bibfnamefont {N.}~\bibnamefont {Bouloufa-Maafa}}, \ and\ \bibinfo
  {author} {\bibfnamefont {E.}~\bibnamefont {Luc-Koenig}},\ }\href {\doibase
  10.1103/PhysRevA.90.033413} {\bibfield  {journal} {\bibinfo  {journal} {Phys.
  Rev. A}\ }\textbf {\bibinfo {volume} {90}},\ \bibinfo {pages} {033413}
  (\bibinfo {year} {2014})}\BibitemShut {NoStop}%
\bibitem [{\citenamefont {Bergeman}\ \emph {et~al.}(2006)\citenamefont
  {Bergeman}, \citenamefont {Qi}, \citenamefont {Wang}, \citenamefont {Huang},
  \citenamefont {Pechkis}, \citenamefont {Eyler}, \citenamefont {Gould},
  \citenamefont {Stwalley}, \citenamefont {Cline}, \citenamefont {Miller},\
  and\ \citenamefont {Heinzen}}]{Bergeman2006}%
  \BibitemOpen
  \bibfield  {author} {\bibinfo {author} {\bibfnamefont {T.}~\bibnamefont
  {Bergeman}}, \bibinfo {author} {\bibfnamefont {J.}~\bibnamefont {Qi}},
  \bibinfo {author} {\bibfnamefont {D.}~\bibnamefont {Wang}}, \bibinfo {author}
  {\bibfnamefont {Y.}~\bibnamefont {Huang}}, \bibinfo {author} {\bibfnamefont
  {H.~K.}\ \bibnamefont {Pechkis}}, \bibinfo {author} {\bibfnamefont {E.~E.}\
  \bibnamefont {Eyler}}, \bibinfo {author} {\bibfnamefont {P.~L.}\ \bibnamefont
  {Gould}}, \bibinfo {author} {\bibfnamefont {W.~C.}\ \bibnamefont {Stwalley}},
  \bibinfo {author} {\bibfnamefont {R.~A.}\ \bibnamefont {Cline}}, \bibinfo
  {author} {\bibfnamefont {J.~D.}\ \bibnamefont {Miller}}, \ and\ \bibinfo
  {author} {\bibfnamefont {D.~J.}\ \bibnamefont {Heinzen}},\ }\href
  {http://stacks.iop.org/0953-4075/39/S813} {\bibfield  {journal} {\bibinfo
  {journal} {J. Phys. B}\ }\textbf {\bibinfo {volume} {39}},\ \bibinfo {pages}
  {S813} (\bibinfo {year} {2006})}\BibitemShut {NoStop}%
\bibitem [{\citenamefont {LeRoy}\ and\ \citenamefont
  {Bernstein}(1970)}]{LeRoy1970}%
  \BibitemOpen
  \bibfield  {author} {\bibinfo {author} {\bibfnamefont {R.~J.}\ \bibnamefont
  {LeRoy}}\ and\ \bibinfo {author} {\bibfnamefont {R.~B.}\ \bibnamefont
  {Bernstein}},\ }\href@noop {} {\bibfield  {journal} {\bibinfo  {journal} {J.
  Chem. Phys.}\ }\textbf {\bibinfo {volume} {52}},\ \bibinfo {pages} {3869}
  (\bibinfo {year} {1970})}\BibitemShut {NoStop}%
\bibitem [{\citenamefont {Stwalley}(1970)}]{Stwalley1970}%
  \BibitemOpen
  \bibfield  {author} {\bibinfo {author} {\bibfnamefont {W.~C.}\ \bibnamefont
  {Stwalley}},\ }\href@noop {} {\bibfield  {journal} {\bibinfo  {journal}
  {Chem. Phys. Lett.}\ }\textbf {\bibinfo {volume} {6}},\ \bibinfo {pages}
  {241} (\bibinfo {year} {1970})}\BibitemShut {NoStop}%
\bibitem [{\citenamefont {Marinescu}\ and\ \citenamefont
  {Sadeghpour}(1999)}]{Marinescu1999}%
  \BibitemOpen
  \bibfield  {author} {\bibinfo {author} {\bibfnamefont {M.}~\bibnamefont
  {Marinescu}}\ and\ \bibinfo {author} {\bibfnamefont {H.~R.}\ \bibnamefont
  {Sadeghpour}},\ }\href {\doibase 10.1103/PhysRevA.59.390} {\bibfield
  {journal} {\bibinfo  {journal} {Phys. Rev. A}\ }\textbf {\bibinfo {volume}
  {59}},\ \bibinfo {pages} {390} (\bibinfo {year} {1999})}\BibitemShut
  {NoStop}%
\bibitem [{\citenamefont {Bussery}\ \emph {et~al.}(1986)\citenamefont
  {Bussery}, \citenamefont {Achkar},\ and\ \citenamefont
  {Aubert-Frecon}}]{Bussery1986}%
  \BibitemOpen
  \bibfield  {author} {\bibinfo {author} {\bibfnamefont {B.}~\bibnamefont
  {Bussery}}, \bibinfo {author} {\bibfnamefont {Y.}~\bibnamefont {Achkar}}, \
  and\ \bibinfo {author} {\bibfnamefont {M.}~\bibnamefont {Aubert-Frecon}},\
  }\href@noop {} {\bibfield  {journal} {\bibinfo  {journal} {Chem. Phys.}\
  }\textbf {\bibinfo {volume} {116}},\ \bibinfo {pages} {319} (\bibinfo {year}
  {1986})}\BibitemShut {NoStop}%
\bibitem [{\citenamefont {Aubert-Fr\'econ}\ \emph {et~al.}(1998)\citenamefont
  {Aubert-Fr\'econ}, \citenamefont {Hadinger}, \citenamefont {Magnier},\ and\
  \citenamefont {Rousseau}}]{Frecon1998}%
  \BibitemOpen
  \bibfield  {author} {\bibinfo {author} {\bibfnamefont {M.}~\bibnamefont
  {Aubert-Fr\'econ}}, \bibinfo {author} {\bibfnamefont {G.}~\bibnamefont
  {Hadinger}}, \bibinfo {author} {\bibfnamefont {S.}~\bibnamefont {Magnier}}, \
  and\ \bibinfo {author} {\bibfnamefont {S.}~\bibnamefont {Rousseau}},\
  }\href@noop {} {\bibfield  {journal} {\bibinfo  {journal} {J. Mol.
  Spectrosc.}\ }\textbf {\bibinfo {volume} {188}},\ \bibinfo {pages} {182}
  (\bibinfo {year} {1998})}\BibitemShut {NoStop}%
\bibitem [{\citenamefont {Comparat}(2004)}]{Comparat2004}%
  \BibitemOpen
  \bibfield  {author} {\bibinfo {author} {\bibfnamefont {D.}~\bibnamefont
  {Comparat}},\ }\href {\doibase http://dx.doi.org/10.1063/1.1626539}
  {\bibfield  {journal} {\bibinfo  {journal} {J. Chem. Phys.}\ }\textbf
  {\bibinfo {volume} {120}},\ \bibinfo {pages} {1318} (\bibinfo {year}
  {2004})}\BibitemShut {NoStop}%
\bibitem [{\citenamefont {Ridinger}\ \emph {et~al.}(2011)\citenamefont
  {Ridinger}, \citenamefont {Chaudhuri}, \citenamefont {Salez}, \citenamefont
  {Fernandes}, \citenamefont {Bouloufa}, \citenamefont {Dulieu}, \citenamefont
  {Salomon},\ and\ \citenamefont {Chevy}}]{Ridinger2011}%
  \BibitemOpen
  \bibfield  {author} {\bibinfo {author} {\bibfnamefont {A.}~\bibnamefont
  {Ridinger}}, \bibinfo {author} {\bibfnamefont {S.}~\bibnamefont {Chaudhuri}},
  \bibinfo {author} {\bibfnamefont {T.}~\bibnamefont {Salez}}, \bibinfo
  {author} {\bibfnamefont {D.~R.}\ \bibnamefont {Fernandes}}, \bibinfo {author}
  {\bibfnamefont {N.}~\bibnamefont {Bouloufa}}, \bibinfo {author}
  {\bibfnamefont {O.}~\bibnamefont {Dulieu}}, \bibinfo {author} {\bibfnamefont
  {C.}~\bibnamefont {Salomon}}, \ and\ \bibinfo {author} {\bibfnamefont
  {F.}~\bibnamefont {Chevy}},\ }\href@noop {} {\bibfield  {journal} {\bibinfo
  {journal} {Europhys. Lett.}\ }\textbf {\bibinfo {volume} {96}},\ \bibinfo
  {pages} {33001} (\bibinfo {year} {2011})}\BibitemShut {NoStop}%
\end{thebibliography}

%

\end{document}